# A NEW PARAMETRIC CLASS OF EXACT SOLUTIONS IN GENERAL RELATIVITY REPRESENTING PERFECT STATIC FLUID BALLS

***B. C. TEWARI****
***MAMTA JOSHI PANT***
***Department of Mathematics, Kumaun University,***
***S.S.J. Campus Almora-263601,India***

***Abstract:*** *We present a new parametric class of spherically symmetric analytic solutions of the general relativistic field equations in canonical coordinates, which corresponds to causal models of perfect fluid balls. These solutions describe perfect fluid balls with infinite central pressure and infinite central density though their ratio is positively finite and less then one. From the solutions of this class we have constructed two causal models in which outmarch of pressure, density is positive and monotonically decreasing and pressure –density ratio is less than one throughout with in the balls. Corresponding to these models we have maximized the Neutron star masses* $3.24M_{\Theta}$ *and* $3.48M_{\Theta}$ *with the linear dimensions 32.09Kms and 34.36Kms respectively with equal surface red shift 0.5811.*

## 1 .Introduction

*Numerous attempts have been reported to obtain parametric classes of exact solutions of Einstein's field equations representing perfect fluid ball in equilibrium.[1]-[17] These solutions have four arbitrary constants. The usual boundary conditions determine three arbitrary constants leaving one undetermined such a solution represents a class of solutions, undetermined constant being a parameter. The undetermined constant to be so adjusted to obtain suitable equation of state i. e. positive and monotonically decreasing expressions for pressure and density, pressure-density ratio must be less then 1 and principle of causality should be obeyed within the ball. In this paper we present yet another parametric class of solutions.*

## 2.Field Equations and Method of Obtaining Analytic Solutions.

**W**e consider the static and spherically symmetric metric in canonical co-ordinates

$$ds^2 = -e^{\lambda} dr^2 - r^2 (d\theta^2 + \sin^2 \theta d\phi^2) + e^{\upsilon} dt^2 \quad (1)$$

where $\lambda$ and $\upsilon$ are functions of $r$ .Einstein's field equations of gravitation for a non empty space- time are

***Email: drbctewari@yahoo.co.in**

$$R_{ij}-\frac{1}{2}Rg_{ij}=-\frac{8\pi G}{c^4}T_{ij} \quad (2)$$

where $R_{ij}$ is a Ricci tensor , $T_{ij}$ is energy- momentum tensor and $R$ the scalar curvature.

The energy–momentum tensor $T_{ij}$ is defined as

$$T_{ij}=(p+\rho c^2)v_i v_j-pg_{ij} \quad (3)$$

where $p$ denotes the pressure distribution, $\rho$ the density distribution and $v_i$ the velocity vector, satisfying the relation

$$g_{ij}v^i v^j=1 \quad (4)$$

Since the field is static, therefore,

$$v^1=v^2=v^3=0 \quad \text{and} \quad v^4=\frac{1}{\sqrt{g_{44}}} \quad (5)$$

Thus we find that for the metric (1) under these conditions the field equation (2) reduces the following:

$$\frac{8\pi G}{c^4}p=e^{-\lambda}(\frac{\upsilon'}{r}+\frac{1}{r^2})-\frac{1}{r^2} \quad (6)$$

$$\frac{8\pi G}{c^4}p=e^{-\lambda}(\frac{\upsilon''}{2}-\frac{\lambda'\upsilon'}{4}+\frac{\upsilon'^2}{4}+\frac{\upsilon'-\lambda'}{2r}) \quad (7)$$

$$\frac{8\pi G}{c^2}\rho=e^{-\lambda}(\frac{\lambda'}{r}-\frac{1}{r^2})+\frac{1}{r^2} \quad (8)$$

where prime $(/)$ denotes differentiation with respect to $r$ . From equations (6) and (7) Tolman [1] obtained following differential equation in $\lambda$ and $\upsilon$ .

$$\frac{d}{dr}(\frac{e^{-\lambda}-1}{r^2})+\frac{d}{dr}(\frac{e^{-\lambda}\upsilon'}{2r})+e^{-\lambda-\upsilon}\frac{d}{dr}(\frac{e^{\upsilon}\upsilon'}{2r})=0 \tag{9}$$

By assigning one of the field variables $\lambda$ and $\upsilon$ as some known functions of $r$ the equation (9) reduces into a form which on integration defines the metric(1) completely and the fluid parameters $p$ and $\rho$ an be calculated from equation (6) and equation (8). Many attempts have been made to solve equation (9) by assuming an *adhoc* relationship between $\upsilon$ and $\lambda$ [2],[4],[9] .

In this paper we use a substitution.

$$e^{\upsilon/2}=U,\ \ V=e^{-\lambda} \tag{10}$$

Thus equation (9) reduces to the following Linear differential equation in *V*.

$$\frac{dV}{dr}-2\left\{\frac{d}{dr}[\ \log\left(\frac{r^3}{rU'+U}\right)\ ]-\frac{2rU}{r^2(rU'+U)}\right\}V=\frac{-2U}{r(rU'+U)} \tag{11}$$

On solving (11) we get,

$$V=e^{-\lambda}=\frac{r^6}{(rU'+U)^2}\left[A-2\int\frac{(rU'+U)Ue^{\int\frac{4Udr}{r(rU'+U)}}dr}{r^7}\right]e^{-\int\frac{4Udr}{r(rU'+U)}} \tag{12}$$

Where $A$ is an arbitrary constant. Our task is to explore the possibilities of choosing $U$ such that the right hand side of equation (12) becomes integrable. In this paper we assume

$$e^{\int\frac{4Udr}{r(rU'+U)}}=r^l(rU'+U)^n \tag{13}$$

$l$ and $n$ are arbitrary constants. Equation (13) results into a second degree homogenous differential equation in U.

$$nr^2U''+(l+2n)rU'+(l-4)U=0 \tag{14}$$

The solution is

$$U = C_1 r^{a+b-1} + C_2 r^{a-b-1} \quad (15)$$

where

$$a = \frac{n-l}{2n}, \qquad b = \frac{1}{2n}\sqrt{(n-l)^2 + 16n}\,, \quad (16)$$

Provided, $n \neq 0$.

$C_1$ and $C_2$ are arbitrary constants. Also(12) is simplified into

$$V = e^{-\lambda} = \frac{r^{8+n-l-(a-b)(n+2)}\left[A - 2I\right]}{\left[(a+b)C_1 r^{2b} + (a-b)C_2\ \right]^{n+2}} \quad (17)$$

where

$$I \equiv \int r^{l-n-9+(a-b)(n+2)}\left[(a+b)C_1 r^{2b} + (a-b)C_2\right]^{n+1}\left[C_1 r^{2b} + C_2\right]dr \quad (18)$$

The solution is complete if (18) is integrated. In the foregoing sections we shall study a method of solving (18) and present a detail discussion of the resulting solution. It may be mentioned here that for $n$=0,we rediscover the class of solutions given by Tolman (Tolman's V solutions).

## 3. New Class of Solutions

The equation(18) can be integrated by the method of substitution, if we assume

$$2b - 1 = l - n - 9 + (a-b)(n+2) \quad (19)$$

In view of (16), the equation (19) yields a quadratic equation in ***l***

$$3(n+1)l^2 + 2\left(n^2 - 11n\right)l + 6n^3 - 13n^2 + 57n = 0 \quad (19a)$$

which solves into

$$l = \frac{-n(n-11) + \sqrt{-n(17n^3 + n^2 + 11n + 171)}}{3(n+1)} \quad , \text{ if } (n+1) \neq 0 \quad (20)$$

$$l = 3.1667 \quad , \text{ if } (n+1) = 0 \quad (20a)$$

Here, we have considered only the positive radical sign ,as corresponding to negative radical sign , $e^{\upsilon}$ becomes singular at the origin. We thus obtain a new class of solutions of equation (10) as follows:

$$e^{\upsilon} = \left( C_1 r^{a+b-1} + C_2 r^{a-b-1} \right)^2 \tag{21}$$

$$e^{-\lambda} = \frac{A r^{-2b}}{\left[(a+b)C_1 r^{2b} + (a-b)C_2\right]^{n+2}} - B\left(C_1^{/} + C_2^{/} r^{-2b}\right) \tag{22}$$

where

$$C_1^{/} = (a+b)(n+2)C_1 \, , C_2^{/} = \left[\ a(n+2) + (n+4)b\ \right]C_2 \tag{23}$$

$$B = \frac{1}{b(a+b)^2 (n+2)(n+3)C_1} \tag{24}$$

We observe that $(e^{\upsilon})_{r=0}$ becomes singular for all values except $-1.3684 \le n \le -1$ and $-0.031 \le n \le 0$. It may be pointed out here that the class of solutions obtained by Wyman M (Wyman 1949) has expression for $e^{\upsilon}$ is similar to (21), however, the two classes of solutions are disjoint.

**4. Properties of the New Class of Solutions**

In view of (21) and (22) we obtain from (6) and (8), the pressure and density distribution respectively.

$$\frac{8\pi G}{c^4} p = \frac{1}{r^2}\left[\begin{array}{l} \left\{ \dfrac{A r^{-2b}}{\left[(a+b)C_1 r^{2b} + (a-b)C_2\right]^{n+2}} - B\left(C_1^{/} + C_2^{/} r^{-2b}\right) \right\} \times \\ \times \left\{ \dfrac{(2a+2b-1)C_1 + (2a-2b-1)C_2 r^{-2b}}{C_1 + C_2 r^{-2b}} \right\} - 1 \end{array}\right] \tag{25}$$

$$\frac{8\pi G}{c^2} \rho = \frac{1}{r^2}\left[\begin{array}{l} \left\{ \dfrac{A\left[(a+b)\{(n+3)2b-1\}C_1 + (2b-1)(a-b)C_2 r^{-2b}\right]}{\left[(a+b)C_1 r^{2b} + (a-b)C_2\right]^{n+3}} \right\} + \\ + (1-2b)BC_2^{/} r^{-2b} + 1 + BC_1^{/} \end{array}\right] \tag{26}$$

In addition to the parameter $n$,the solutions (21) and(22)contain three arbitrary constants $C_1$, $C_2$ and $A$. These are to be determined by matching the solutions (21) and (22) with Schwarzschild exterior solution for a ball of mass $M$ and linear dimension $2r_b$:

$$p(r_b) = 0 \tag{27}$$

$$e^{-\lambda(r_b)} = 1 - 2u \tag{28}$$

$$e^{\upsilon(r_b)} = 1 - 2u \tag{29}$$

where,

$$u = \frac{GM}{c^2 r_b} \tag{30}$$

Consequently,

$$C_1 = \frac{[(2a - 2b - 1)u - a + b + 1]}{2b\left(\sqrt{1-2u}\right)r_b^{\,a+b-1}} \tag{31}$$

$$C_2 = -\frac{[(2a + 2b - 1)\,u - a - b + 1]}{2b\left(\sqrt{1-2u}\right)\,r_b^{\,a+b-1}} \tag{32}$$

$$A = \left[1 - 2u + \left(C_1^{\prime} + C_2^{\prime} r_b^{-2b}\right)\right]\left[(a+b)C_1 r_b^{2b} + (a-b)C_2\right]^{n+2} r_b^{\,2} \tag{33}$$

For $e^{\upsilon}$ to be definitely positive in the region $0 \le r \le r_b$, we must have $C_1\ , C_2 \rangle 0$ .Thus in view of (31) and (32) we have;

$$\frac{a+b-1}{2a+2b-1} \le u \le \frac{a-b-1}{2a-2b-1} \tag{34}$$

The central values of pressure and density are infinite, however, the limiting value of their ratio is finite and equal to the limiting value of $\frac{dp}{d\rho}$:

$$\left(\frac{p}{\rho c^2}\right)_{r\to 0} = \frac{1}{c^2}\left(\frac{dp}{d\rho}\right)_{r\to 0} = -\frac{[2a + 2b - 1 + b(n+3)(a+b)]}{[1 + b(n+3)(a+b)]} \tag{35}$$

It has been calculated that for values of $n$ in the interval [-1.3684, -1.058), the right hand side of (35) is negative thus making adiabatic sound speed imaginary and corresponding to the values of $n$ in the interval [-1.058, -1], the right hand side of (35) is positive and less than 1 i.e. causality principle is obeyed at the centre ( Table- I). Hence for meaningful solutions $n$ ranges within the interval [-1.058, -1].

Variation of central pressure- central density ratio / square of adiabatic sound speed at the centre and parameters of indices of $e^{\upsilon}$ for different values of $n$

Table - I

| S. N. | $n$ | $L$ | $a$ | $b$ | $\left(\frac{p}{\rho c^2}\right)_{r\to 0} = \frac{1}{c^2}\left(\frac{dp}{d\rho}\right)_{r\to 0}$ |
|---|---|---|---|---|---|
| 1 | -1 | 3.1667 | 2.08335 | -0.58339 | 0.333 |
| 2 | -1.01 | 3.1870 | 2.07773 | -0.59714 | 0.2656 |
| 3 | -1.02 | 3.2074 | 2.07228 | -0.61058 | 0.2037 |
| 4 | -1.03 | 3.228 | 2.06699 | -0.62367 | 0.1465 |
| 5 | -1.04 | 3.2486 | 2.06184 | -0.63644 | 0.0892 |
| 6 | -1.05 | 3.2693 | 2.05680 | -0.64890 | 0.04388 |
| 7 | -1.058 | 3.2860 | 2.05294 | -0.65869 | 0.00749 |

## 5. Particular Members of Class

In this section we shall present a detail study of the particular solutions corresponding to $n$ = -1 and $n$ = -1.05.

(I) for $n$ = -1, the solution is

$$e^{\upsilon} = \left( C_1 r^{0.5} + C_2 r^{1.667} \right)^2$$

$$e^{-\lambda} = \frac{A r^{1.17}}{\left[1.5 C_1 r^{-1.17} + 2.667 C_2\right]} + \frac{0.38}{C_1}\left(1.5 C_1 + 0.33 C_2\ r^{1.17}\right)$$

$$\frac{8\pi G}{c^4} p =$$

$$\left[\left\{\frac{A r^{1.17}}{\left[1.5C_1 r^{-1.17} + 2.667C_2\right]} + \frac{0.38}{C_1}\left(1.5C_1 + 0.33C_2\, r^{1.17}\right)\right\}\frac{\left(2C_1 + 4.33C_2\, r^{1.17}\right)}{\left(C_1 + C_2\, r^{1.17}\right)} - 1\right]\frac{1}{r^2}$$

$$\frac{8\pi G}{c^2}\rho = \left[\left\{\frac{\left(-5.0004C_1 - 5.7788C_2\, r^{1.17}\right)A}{\left[1.5C_1 r^{-1.17} + 2.667C_2\right]^2}\right\} - \frac{0.272C_2\, r^{1.17}}{C_1} + 0.43\right]\frac{1}{r^2}$$

In view of (34) $e^{\upsilon}$ to be definitely positive in the region $0 \le r \le r_b$, then

$$0.25 \le u \le 0.3846$$

Corresponding to $u = 0.3$ and in view of (31), (32) and (33) the constants are

$$C_1 r_b^{0.5} = 0.497,\ C_2 r_b^{1.667} = 0.1355,\ A r_b^{2.84} = -0.2260$$

In Table II the march of pressure, density, pressure – density ratio and square of adiabatic sound speed $\frac{dp}{d\rho}$ is given for $u = 0.3$. We observe that pressure and density decrease monotonically with the increase of radial coordinate, pressure – density ratio and square of adiabatic sound speed is positive and less than *1* throughout within the ball.

**Table II**

The march of pressure, density, pressure – density ratio and square of adiabatic sound speed within the ball corresponding to *n* = -1 with $u = 0.3$.

| $r/r_b$ | $\frac{8\pi G}{c^4} p r_b^2$ | $\frac{8\pi G}{c^2}\rho\, r_b^2$ | $\frac{p}{\rho c^2}$ | $\frac{1}{c^2}\left(\frac{dp}{d\rho}\right)$ |
|---|---|---|---|---|
| 0 | $\infty$ | $\infty$ | 0.333 | 0.333 |
| 0.1 | 16.6018 | 42.9419 | 0.3866 | 0.3708 |
| 0.2 | 4.7525 | 10.9999 | 0.4320 | 0.4215 |
| 0.3 | 2.2907 | 5.1544 | 0.4444 | 0.4801 |
| 0.4 | 1.3257 | 3.1322 | 0.4232 | 0.5300 |
| 0.5 | 0.8273 | 2.2045 | 0.3752 | 0.5960 |
| 0.6 | 0.5269 | 1.7025 | 0.3094 | 0.6450 |
| 0.7 | 0.3204 | 1.3990 | 0.2290 | 0.6933 |
| 0.8 | 0.1859 | 1.1999 | 0.1549 | 0.8146 |
| 0.9 | 0.0807 | 1.0609 | 0.07560 | 0.7905 |
| 1.0 | 0 | 0.9588 | 0 | 0.7439 |

(II) For $n$ = -1.05, the solution is

$$e^{\upsilon} = \left( C_1 r^{0.4078} + C_2 r^{1.7057} \right)^2$$

$$e^{-\lambda} = \frac{A r^{1.2978}}{\left[1.4078 C_1 r^{-1.2978} + 2.7057 C_2\right]^{0.95}} + \frac{0.4197}{C_1}\left(1.3375 C_1 + 0.0397 C_2\, r^{1.2978}\right)$$

$$\frac{8\pi G}{c^4} p =$$

$$\left[ \left\{ \frac{A r^{1.2978}}{\left[1.4078 C_1 r^{-1.2978} + 2.7057 C_2\right]^{0.95}} + \frac{0.4197}{C_1}\left(1.3375 C_1 + 0.0397 C_2\, r^{1.2978}\right) \right\} \times \frac{\left(1.8175 C_1 + 4.411 C_2\, r^{1.2978}\right)}{\left(C_1 + C_2\, r^{1.2978}\right)} - 1 \right] \frac{1}{r^2}$$

$$\frac{8\pi G}{c^2}\rho = \left[ \left\{ \frac{\left(-4.97 C_1 - 6.217 C_2\, r^{1.2978}\right) A}{\left[1.4078 C_1 r^{-1.2978} + 2.7057 C_2\right]^{1.95}} \right\} - \frac{0.0383 C_2\, r^{1.2978}}{C_1} + 0.439 \right] \frac{1}{r^2}$$

In view of (34) $e^{\upsilon}$ to be definitely positive in the region $0 \le r \le r_b$ ,then

$$0.2246 \le u \le 0.3866$$

Corresponding to $u = 0.3$ and in view of (31), (32) and (33) the constants are

$$C_1 r_b^{0.4078} = 0.4657,\ C_2 r_b^{1.7057} = 0.1667,\ A r_b^{2.92} = -0.2446$$

In Table III the march of $p, \rho, \frac{p}{\rho c^2}$ $and \frac{dp}{c^2 d\rho}$ within the fluid ball for $u = 0.3$. We observe that pressure and density decrease monotonically with increase of radial coordinate $\frac{p}{\rho c^2}$ $and$ $\frac{1}{c^2}\left(\frac{dp}{d\rho}\right)$ and less than 1 throughout within the ball .

Table III

| $r/r_b$ | $\frac{8\pi G}{c^4} p r_b^2$ | $\frac{8\pi G}{c^2} \rho\ r_b^2$ | $\frac{p}{\rho c^2}$ | $\frac{1}{c^2}\left(\frac{dp}{d\rho}\right)$ |
|---|---|---|---|---|
| 0 | $\infty$ | $\infty$ | 0.04388 | 0.04388 |
| 0.1 | 4.9136 | 44.2266 | 0.1111 | 0.0854 |
| 0.2 | 2.11629 | 11.4429 | 0.1849 | 0.1370 |
| 0.3 | 1.2867 | 5.3841 | 0.2389 | 0.2059 |
| 0.4 | 0.86436 | 3.3462 | 0.2583 | 0.2842 |
| 0.5 | 0.59813 | 2.3905 | 0.2502 | 0.3461 |
| 0.6 | 0.41072 | 1.8728 | 0.2193 | 0.4375 |
| 0.7 | 0.27033 | 1.5589 | 0.1734 | 0.5238 |
| 0.8 | 0.16126 | 1.3459 | 0.1198 | 0.6428 |
| 0.9 | 0.07082 | 1.2068 | 0.0586 | 0.6930 |
| 1.0 | 0 | 1.09915 | 0 | 0.7350 |

In Table IV, the variation of surface density ($\rho_b$), maximum Neutron star mass (multiple of solar mass $M_\Theta$), linear dimension corresponding to $u = 0.3$ with constant surface red shift $Z_b = [(1 - 2u)^{-0.5} - 1] \approx 0.5811$ and different values of $n$ .

Table IV

| S. N. | $n$ | $\frac{8\pi G}{c^2} \rho_b r_b^2$ | $M/M_\Theta$ | $\approx 2 r_b$ Kms | $Z_b$ |
|---|---|---|---|---|---|
| 1 | -1 | 0.9588 | 3.24 | 32.09 | 0.5811 |
| 2 | -1.01 | 0.9918 | 3.31 | 32.64 | 0.5811 |
| 3 | -1.02 | 1.029 | 3.36 | 33.26 | 0.5811 |
| 4 | -1.03 | 1.0435 | 3.39 | 33.48 | 0.5811 |
| 5 | -1.04 | 1.079 | 3.44 | 34.05 | 0.5811 |
| 6 | -1.05 | 1.099 | 3.48 | 34.36 | 0.5811 |
| 7 | -1.058 | 1.1372 | 3.52 | 34.69 | 0.5811 |

We observe that by decreasing the values of $n$ the surface density, the mass and the linear dimension of Neutron star are increasing in nature.

## 6. Conclusion

For meaningful realistic fluid ball pressure and density must be positive and monotonically decreasing with the increase of radial coordinate r, $\frac{p}{\rho c^2}$ *and* $\frac{1}{c^2}\left(\frac{dp}{d\rho}\right)$ must be positive and less than 1 throughout within the ball. For meaningful solution $e^{\upsilon} = \left( C_1 r^{\alpha} + C_2 r^{\beta} \right)^2$, where $\alpha$ and $\beta$ should be positive so that $e^{\upsilon}$ at centre becomes non-singular (positive and finite) but converse is not true.

### References

[1] Tolman, R. C.: Phy. Rev. **55**, 367(1939)

[2] Wyman, M.: Phy. Rev. **75**, 1930(1949)

[3] Buchdahl,H.: Astrophys. J.**140**,1512, (1964)

[4] B. Kuchowocz: Acta. Phys. Polan **33,** 541 (1968)

[5] Leibovitz, C: Phy. Rev. D **185**, 1664(1969)

[6] Adler R,: J .Math.Phys.**15,**727 ,(1974)

[7] Matese, J.J. and Whitman, P.G.:, Phys. Rev. D **22**,1270(1980)

[8] Pant, D.N. and Sah, A. : Phys. Rev. D **26**,1254(1982)

[9] Pant, D.N. and Sah, A. : Phys. Rev. D **32**,1358(1985)

[10] Pant, D.N. and Pant, Neeraj : Jour. Math. Phys. **34**, 2440(1993)

[11] Pant, D.N.: Astro. Space Sci. **215**, 97 (1994)

[12] Pant, Neeraj : Astro. Space Sci. **240**, 187(1996)

[13] Rosquist,k: Phy. Rev. D **59,**044022 (1999)

[14] Dev K and Gleiser M : Gen.Rel.Grav.**34**, 1793, (2002 )

[15] Maharaja S D and Chaisi M : Gen.Rel.Grav.**38**, 1723(2006)

[16] Walter Simon : Gen Rel. Grav**.40**,2591(2008)

[17] S Thirukkanesh and Maharaj Class. Q .Gravity. **25**,235001,(2008)